\documentclass[aps,prl,twocolumn,superscriptaddress]{revtex4}
\usepackage{graphicx,epsf}
\usepackage{amsmath}

\newcommand{\CeAu}{CeCu$_{6-x}$Au$_x$}

\newcommand{\YRS}{YbRh$_2$Si$_2$}

\begin{document}

\title{
Zeeman-driven Lifshitz transition:\\
A scenario for the Fermi-surface reconstruction in \YRS
}

\author{Andreas Hackl}
\affiliation{Department of Physics, California Institute of Technology, Pasadena, CA 91125, USA}
\author{Matthias Vojta}
\affiliation{Institut f\"ur Theoretische Physik, Technische Universit\"at Dresden,
01062 Dresden, Germany}

\date{\today}

\begin{abstract}
The heavy-fermion metal \YRS\ displays a field-driven quantum phase transition where
signatures of a Fermi-surface reconstruction have been identified, often interpreted as
breakdown of the Kondo effect. We argue that instead many properties of the material can
be consistently described assuming a Zeeman-driven Lifshitz transition of narrow heavy-fermion
bands. Using a suitable quasiparticle model, we find a smeared jump in the Hall
constant and lines of maxima in susceptibility and specific heat, very similar to experimental
data. An intermediate non-Fermi liquid regime emerges due to the small effective Fermi
energy near the transition. Further experiments to discriminate the different scenarios
are proposed.
\end{abstract}

\maketitle


Quantum criticality in heavy-fermion metals constitutes an exciting and active research area
\cite{geg08,hvl}. Among the various compounds investigated, \CeAu\
and \YRS\ stand out as possible candidates for the realization of a Kondo-breakdown
quantum phase transition (QPT). Both display a transition from a non-magnetic to an
antiferromagnetic (AF) metallic phase, with critical behavior which appears inconsistent
with the predictions of the Landau-Ginzburg-Wilson (LGW) theory of
three-dimensional (3d) AF criticality. This has prompted speculations about a novel class
of QPT where the Kondo effect, responsible for the formation of heavy
quasiparticles (QP), itself becomes critical, implying a reconstruction of the entire Fermi
surface across the transition \cite{coleman01,si01,flst1}.

\YRS\ \cite{trovarelli00}, where the QPT is tuned by an applied magnetic field
\cite{gegenwart02}, is particularly interesting, because signatures of a Fermi-surface
reconstruction have been identified in the Hall effect \cite{paschen04,friede10}. Those
measurements allowed to trace a line $T_{\rm Hall}(B)$ in the temperature--field phase
diagram where the crossover between two Fermi-surface configurations is supposed to
occur, with the line terminating at the QPT at $B_c$ and the crossover width $\Delta
B(T)$ being approximately linear in temperature \cite{friede10}. In addition, distinct
changes in thermodynamic quantities have been employed to map out other crossover lines:
magnetic susceptibility ($\chi$), magnetization, and magnetostriction data result in a
$T^\ast(B)$ line which roughly matches the $T_{\rm Hall}(B)$ line
\cite{gegenwart06,gegenwart07}, while maxima in the specific heat coefficient, $\gamma =
C/T$, lead to a lower $T_{\rm max}(B)$ line, also terminating at the QPT at $B_c$
\cite{oeschler08}. Low-$T$ Fermi-liquid behavior, identified by a $T^2$ behavior of the
resistivity, sets in at an even lower $T_{\rm FL}(B)$.

While the crossovers near the $T^\ast$ and $T_{\rm Hall}$ lines are frequently assumed to
be signatures of Kondo-breakdown quantum criticality, there are features which are
not easily consistent with this hypothesis:
(i) Under both isoelectronic doping and pressure, the magnetic phase boundary was found
to move, while the $T^\ast$ line moved very little \cite{friede09,custers10}.
Assuming that (chemical) pressure tunes the competition between Kondo screening and
non-local interactions, the behavior of the $T^\ast$ line is rather unexpected.
(ii) Signatures of the $T^\ast$ crossovers are never observed in zero field.
(iii) In a $T$ range between 0.1 and 1.5\,K, $T^\ast = g^\ast \mu_B B$
\cite{kb_foot} within error bars, with $g^\ast \approx 4.7$ for $B$ applied in the a-b
plane. In the same $T$ range, $T_{\rm max} = g_{\rm max} \mu_B B$ with $g_{\rm max}
\approx 1.7$. This points towards Zeeman physics as a key player.
%
%
(iv) In a quantum critical scenario, the $T$-linear width of the Hall crossover
\cite{friede10} suggests $\nu z=1$, where $\nu$ and $z$ are the correlation length and
dynamical critical exponents. This is at odds with $\nu z\approx 0.7$ inferred from
measurements of the Gr\"uneisen parameters $\Gamma_p$ \cite{kuechler03}
and $\Gamma_H$ \cite{tokiwa09}.
(v) Even at $B=B_c$, distinct low-temperature crossovers are seen in $C(T)$
\cite{gegenwart02,oeschler08}, $\Gamma_{p,H}(T)$ \cite{kuechler03,tokiwa09},
and the thermopower \cite{hartmann10}.
Thus, quantum criticality is at best restricted to very small $T$.
%
Finally, we note that no theoretical
modeling of the behavior near $T_{\rm Hall}$, $T^\ast$ is available to date.

In this Letter, we propose an alternative explanation for the intriguing crossovers
in \YRS. Our scenario is that of a Fermi-surface reconstruction of a 
narrow heavy-fermion band (or piece thereof) via one or more Lifshitz transitions driven by
Zeeman splitting \cite{kusm08,pw_esr}.
We shall show that this scenario naturally explains many elevated-temperature features,
such as maxima in $\chi$ and $\gamma$ at $T^\ast,T_{\rm max} \propto B$ and a Hall
crossover with a $T$-linear width. Importantly, these properties do {\em not} emerge in a
quantum critical regime, but rather in a ``high-temperature'' regime above a tiny
intrinsic energy scale (the depth of the assumed Fermi pockets). This regime is
characterized by apparent non-Fermi liquid behavior, similar to the experimental data.
Our scenario leads to a variety of predictions, to be tested in future experiments.

\begin{figure}[!t]
\epsfxsize=3.4in
\centerline{\epsffile{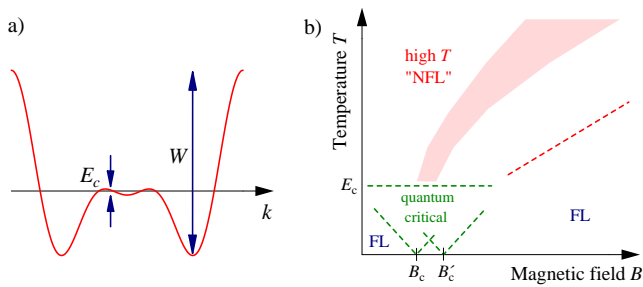}}
\caption{
a) Assumed band structure (schematic) of the heavy-fermion QP,
with shallow Fermi pockets which induce a large zero-field DOS and lead to
one (or more) field-induced Lifshitz transitions at small fields $B_c$, with $g \mu_B B_c \sim E_c$.
b) Temperature--field phase diagram (schematic), with a quantum critical regime
associated with the Lifshitz transitions at $B_c, B_c'$; this regime is restricted to
temperatures $T\ll E_c$ due to the small effective bandwidth $E_c$.
The elevated-temperature regime, $T\gg E_c$, is characterized by apparent non-Fermi liquid
behavior, with distinct crossovers occurring in the shaded region of the phase diagram,
where the thermodynamics is that of a Zeeman-tuned Schottky anomaly.
}
\label{fig:schem}
\end{figure}


{\it Band structure and Lifshitz transition.}
We shall assume that most of the physics of \YRS\ at temperatures and fields below the
Kondo scale, $T\ll T_0 \approx 20$\,K and $B\ll B_0\approx 10$\,T, can be described in
terms of heavy-fermion QP, implying that Kondo screening is effective near
the $T^\ast$ line \cite{pw_esr}.
%

Consider a band structure with an overall bandwidth $W$ and a narrow piece of band at the
Fermi level, the latter with a tiny effective bandwidth $E_c \ll W$ and a correspondingly
small velocity (Fig.~\ref{fig:schem}a). As a result, the zero-field density of states
(DOS) displays a pronounced peak at the Fermi level, originating from Fermi pockets with
an effective Fermi energy $\lesssim E_c$.
Then, a tiny Zeeman field $g \mu_B B \sim E_c$, where $g$ denotes the QP $g$
factor, will split the bands such that the Fermi pockets disappear via zero-temperature
Lifshitz transitions. At low temperatures, $T\ll E_c$,
a standard quantum critical regime \cite{ssbook} near a critical field $B_c$ marks the
crossover between two Fermi-liquid regimes at small and large fields, the crossover being
associated with a Fermi-surface reconstruction. (For a band structure as in
Fig.~\ref{fig:schem}a, there will be {\em two} separate Lifshitz transitions for the two
spin species. Provided that the two critical fields are close, only measurements at
ultralow temperatures, $T\ll g\mu_B|B_c-B_c'|$, will resolve two transitions.)

Most importantly, the Lifshitz-transition signatures in the $T$--$B$ plane extend to
temperatures much larger than $E_c$, because, as the DOS peak is split, the
thermodynamics can be understood in terms of a Zeeman-tuned Schottky anomaly. As we shall
demonstrate, quantities like $\chi(T)$ and $\gamma(T)$ generically display maxima at
temperatures which follow $T\propto B$ for $T\gg E_c$.
%
Moreover, interesting violations of Fermi-liquid behavior as function of $T$ are found
for temperatures above the scale $B_c$. Transport properties are similarly sensitive to
the Lifshitz transitions and associated crossovers, and we will discuss the Hall effect
below.

The results of this theoretical scenario match salient properties of \YRS, assuming that
$E_c \sim 50$\,mK, $B_c \sim 50$\,mT (for $B$ in the a-b plane), while $W\sim
10-20$\,K is of order of the Kondo scale \cite{kb_foot,zwick_note}.


{\it Thermodynamics.}
For weakly interacting QP, thermodynamic quantities are entirely determined
by the QP DOS $\rho(E)$.
As elucidated above, the low-temperature and high-temperature regimes,
characterized by $T\ll E_c$ and $E_c\ll T\ll W$ need to be distinguished.
(The regime of $T\gtrsim W$ shall not be of interest here, as QP physics is
not applicable for $T\gtrsim T_K$.)

For low $T$ near a Zeeman-driven Lifshitz transition at $B_c$, analytical
results can be easily derived \cite{ssbook}.
Outside the critical regime, i.e., for $T\ll g \mu_B |B-B_c|$, the behavior is
Fermi-liquid like, and the magnetic Gr\"uneisen parameter (or magnetocaloric effect) is a
constant reflecting the difference between the DOS of the two spin species, $\Gamma_H =
-(dM/dT)/C_B \propto \rho_\uparrow - \rho_\downarrow$.
In contrast, in the quantum critical regime, $g \mu_B |B-B_c|\ll T \ll E_c$, the free
energy follows $F\propto T^{1+d/2}$ for a band-edge singularity in $d$ space dimensions.
The critical piece of the specific heat is then $C_{\rm cr}\propto T^{d/2}$, similar to
that of the magnetization, $M_{\rm cr}\propto T^{d/2}$. Taking into account the
Fermi-liquid background contribution, this results in $\Gamma_H \propto
T^{d/2-2}$, with the sign depending on the type of Lifshitz transition: If pockets
disappear (appear) with increasing field, then $\Gamma_H > 0$ ($\Gamma_H <0$). If the
quantum critical regimes of two nearby Lifshitz transitions overlap
(Fig.~\ref{fig:schem}b), then the critical contributions to thermodynamic quantities
simply add, without further qualitative changes.
For \YRS\ these results apply to $T\ll 50$\, mK.

More relevant is the high-temperature regime, which -- upon tuning $B$ -- will be
dominated by the competition between the two scales $B$ and $T$, resulting in crossovers
as function of $g \mu_B B/T$. Concrete results in this regime depend on the DOS on all
energies up to ${\rm max}(g\mu_B B,T)$, but some effects can be illustrated in the narrow-band
limit, i.e., in a toy model of a Zeeman-split local fermionic level at $\pm h$ with $h= g
\mu_B B/2$. For $T\gg h$, $\chi \propto 1/T$. For
fixed $h$, $\chi(T)$ has a maximum at $T^\ast/h=0.65$ whereas $\gamma(T)$ has a maximum
at $T_{\rm max}/h=0.31$ \cite{max_foot}.

\begin{figure}[!t]
\epsfxsize=3.5in
\centerline{\epsffile{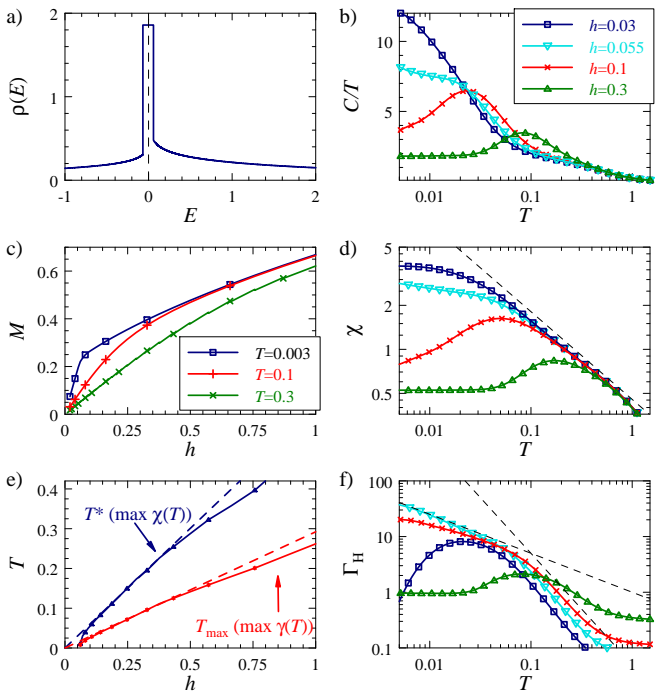}}
\vspace*{-12pt}
\caption{
Thermodynamic properties near a Zeeman-driven Lifshitz transition.
a) Input DOS $\rho(E)$ with bandwidth $W=4$ and a low-energy peak of width $E_c=0.12$.
Step discontinuities are crossed at fields $h_c = 0.055$ and $h'_c = 0.08$, $h= g\mu_B B/2$.
b) Specific heat coefficient $\gamma(T)$ for different applied fields.
c) Magnetization $M(h)$, showing a pronounced kink at a temperature-dependent field.
d) Magnetic susceptibility $\chi(T)$; the dashed line is a power law $\chi\propto
T^{-0.6}$.
e) Location of maxima in $\chi(T)$ and $\gamma(T)$ in the $T$--$h$ phase diagram; the
dashed lines are linear fits through the origin.
f) Magnetocaloric effect $\Gamma_H(T)$; the dashed lines show power laws $\Gamma_H\propto
T^{-0.7}$ and $T^{-2}$.
}
\label{fig:td1}
\end{figure}

A full set of numerical results is presented in Fig.~\ref{fig:td1}, for free fermions
with the sample DOS shown in Fig.~\ref{fig:td1}a. This DOS has two step discontinuities
(e.g. from band edges in 2d), leading to two nearby Lifshitz transitions at
$h_c$, $h'_c$. While Fermi-liquid laws are obeyed at asymptotically low $T$ at all fields
away from the Lifshitz transitions, they are generically violated at elevated $T$.
In particular, near $h_c$, $\gamma(T)$ shows a weak apparent divergence for $T>0.1$, with
a pronounced upturn for $T<0.1$ and saturation only for $T<0.005$, while $\chi(T)$
follows an approximate power law $T^{-0.6}$ down to $T=0.05$ before it saturates. A field
cuts off the apparent low-temperature singularities, leading to maxima in $\gamma(T)$ and
$\chi(T)$ for $h>h_c,h'_c$. These maxima follow $T^\ast,T_{\rm max}\propto h$ with
$T^\ast/T_{\rm max}\approx 2.1$ over an intermediate range of fields (physics of a
Schottky anomaly), but the non-zero background DOS induces a curvature which becomes
significant at large fields.
The magnetocaloric effect is positive; near $h_c$ it displays a crossover between two
apparent power laws, $\Gamma_H\propto T^{-0.7}$ for $T<0.05$ and $\Gamma_H\propto T^{-2}$
for $0.1<T<0.7$ -- the latter strong divergence is rooted in the apparent divergence of
$\chi(T)$, together with the weak $T$ dependence of $\gamma$. (The critical
$\Gamma_H\propto 1/T$ is seen at $h_c$, $h'_c$ only for $T<0.005$.)
It should be emphasized that none of the power laws above $T=0.01$ is of asymptotic character;
they rather represent crossover behavior arising from the interplay of peak and
background DOS. Consequently, the exponents are non-universal.
We note that the precise {\em shape} of the DOS peak is unimportant for $T\gtrsim E_c$,
in particular, 3d Lifshitz transitions yield qualitatively similar results.


{\it Transport and Hall effect.}
In contrast to thermodynamics, transport properties depend on the actual band structure.
In the absence of detailed experimental or ab-initio information \cite{zwick_note}, we
focus on the shape and width of the crossover in the Hall constant $R_H$ at a generic
Zeeman-driven Lifshitz transition.
Being interested in $T,B\ll T_K$, we shall employ a Boltzmann approach with
momentum-independent QP scattering rate \cite{scatt_note}. In this approximation, the
critical piece follows $R_{H,\rm cr} \propto (h_c-h)^{d/2}$ at $T=0$ (while non-critical
Zeeman-induced changes in $R_H$ are linear in $h$).

\begin{figure}[!t]
\epsfxsize=3.5in
\centerline{\epsffile{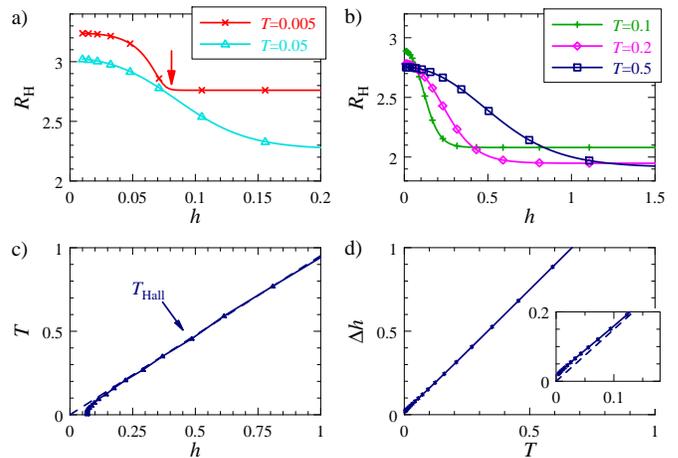}}
\vspace*{-12pt}
\caption{
Hall coefficient of a square-lattice tight-binding model with hopping 0.05 and filling
1.7. The Zeeman-driven Lifshitz transition is at $h_c=0.08$.
a,b) $R_H(h)$ for different temperatures. The sharp crossover at low $T$ arises from the
Lifshitz transition (arrow). c) Location of the Hall crossover in the $T$--$h$ phase diagram,
defined by the maximum location of $|dR_H/dh|$. d) Full width at half maximum of $|dR_H/dh|$.
} \label{fig:hall1}
\end{figure}

Sample results for the Hall constant are shown in Fig.~\ref{fig:hall1} for a
square-lattice tight-binding model. $R_H$ approaches constant values in the limits of
small and large fields, with a pronounced crossover in between. For $T\to0$, this
``smeared step'' can be understood as follows: On the high-field side, there is a kink at
the Lifshitz transition (Fig.~\ref{fig:hall1}a), while the behavior on the low-field side
arises from deviations from parabolic dispersion. The location and width of the crossover
peak in $dR_H/dh$ at fixed $T$ allow to extract a crossover line $T_{\rm Hall}(h)$ and
width $\Delta h(T)$, plotted in Figs.~\ref{fig:hall1}c and d. (Note that the peak
position in $dR_H/dh$ for $T\to 0$ does not coincide with the precise location of the
Lifshitz transition.)
The behavior of $T_{\rm Hall}(h)$ is similar to that of
$T^\ast$ and $T_{\rm max}$, with a large-field slope of $T_{\rm Hall}/h \approx 0.98$
\cite{thall_note}. The crossover width varies linearly with $T$ with a small offset of
$\Delta h(T\to0) \approx h_c/4$. This saturation signals that the $T=0$ behavior of $R_H(h)$ is
continuous. Importantly, the $T$-linear width is a result of thermal smearing, not of
collective effects. We have verified that the qualitative results are generic for Lifshitz
transitions in both 2d and 3d (up to a non-critical background in $R_H$ arising from
other bands).
In application to \YRS, our results imply that the crossover width arising from a Lifshitz
transition can well ``look'' $T$-linear down to 30\,mT.


{\it Discussion.}
So far, collective instabilities of the heavy QP are not included in the description.
They can be expected to be important at low $T$:
Plugging the large-field $g$ factor of 3.6 into the QP calculation, the field-induced
energy and temperature scales are too small by a factor of 3--4, as seen in the slopes of
$T^\ast$ and $T_{\rm max}$. A plausible explanation is a large enhancement of the
low-field $g$ factor due to incipient ferromagnetism. Indeed, the QP Zeeman splitting
can be strongly enhanced by ferromagnetic correlations \cite{pw09}, known to be present
in \YRS\ \cite{gegenwart05}.

Which experiments can reliably distinguish between the scenarios of
(i) a Kondo-breakdown transition and
(ii) a Zeeman-driven Lifshitz transition
as source of the Fermi-surface reconstruction?
The principal distinction is that in (i) the low-field phase in the absence of
magnetism is expected to be a true non-Fermi liquid, i.e., a metallic spin liquid, most
likely of the fractionalized Fermi-liquid type \cite{flst1}. In contrast, in (ii) all
phases have Fermi-liquid character in the low-temperature limit, with $\gamma\to{\rm
const}$ and the Wiedemann-Franz law being satisfied. Experiments on Ir-doped or Ge-doped
\YRS\ \cite{friede09,custers10}, where magnetism is suppressed down to very low $T$ and
$B$, can access this region of the phase diagram.
The practical problem is that probing the true low-$T$ behavior requires temperatures
significantly below 50 mK. Existing data do not cover this regime in a sufficient manner.

A further distinction is in the response of the phase boundary and crossover lines to
changes in system parameters, such as doping and pressure. Scenario (i) implies that the
$T^\ast$ line arises from the competition between Kondo screening and inter-moment
interactions, and thus should be sensitive to changes in the hybridization strength
induced by pressure. In contrast, in scenario (ii) the phase boundary will be more robust
and only react to strong changes in hybridization matrix elements or in the electron
concentration. Existing data, where (chemical) pressure has little influence on the
$T^\ast$ line \cite{friede09,custers10}, appear more consistent with (ii). Thus, we
propose to study \YRS\ using dopants with different valence. Here, an influence on the
phase boundary can be expected for sizeable doping levels.

Our scenario yields a connection between incipient ferromagnetism and the slopes of the
$T^\ast$ and $T_{\rm max}$ lines. If certain dopants enhance (reduce) the ferromagnetic
tendencies, then those slopes should increase (decrease).

Finally, we touch upon the role of AF. This is an instability independent of the Lifshitz
transition, but the two may influence each other depending on Fermi-surface details.
Fluctuations of AF order will influence both thermodynamics and transport at low $T$ near
the AF transition, e.g., they will modify the upturn in $\gamma(T)$. Inside the AF phase,
it is conceivable that strong ordering will modify the band structure such that the
shallow Fermi pockets are removed from the Fermi level. Hence, we predict that the
low-$T$ part of the $T^\ast$ line will be smeared in sufficiently Co-doped \YRS.


{\it Summary.}
We have shown that key features of the field-driven QPT in \YRS\ can be consistently
explained in the framework of Zeeman-driven Lifshitz transition, with Kondo screening
remaining intact. Zeeman splitting of shallow Fermi pockets causes anomalies in both
thermodynamic and transport properties, including apparent non-Fermi liquid behavior.
Most remarkable is a smeared jump in the Hall constant upon variation of the field which
signals a Fermi surface reconstruction, with a crossover width proportional to $T$ over a
wide range of temperatures.
While it is clear that collective effects are relevant at very low temperatures, our
results suggest that a large portion of the \YRS\ phase diagram can be understood in
terms of QP Lifshitz physics.



We enjoyed extensive exchange with M. Brando, P. Gegenwart, and O. Stockert, who
suggested the relevance of Zeeman physics in \YRS. We further thank S. Friedemann, A.
Rosch, F. Steglich, and P. W\"olfle for discussions. This research was supported by DFG
FOR 960.



\vspace*{-15pt}
\begin{thebibliography}{}
\vspace*{-15pt}

\bibitem{geg08}
P. Gegenwart {\em et al.},
Nature Phys. {\bf 4}, 186 (2008).

\bibitem{hvl}
H. v. L\"ohneysen {\em et al.},
Rev. Mod. Phys. {\bf 79}, 1015 (2007).

\bibitem{coleman01}
P.~Coleman, C.~P\'epin, Q.~Si, and R.~Ramazashvili,
J. Phys: Condens. Matt. {\bf 13}, R723, (2001).

\bibitem{si01}
Q. Si {\em et al.},
Nature (London) \textbf{413}, 804 (2001).
%

\bibitem{flst1}
T. Senthil {\em et al.},
Phys. Rev. Lett. \textbf{69}, 216403 (2003).

\bibitem{trovarelli00}
O. Trovarelli {\em et al.},
Phys. Rev. Lett. {\bf 85}, 626 (2000).

\bibitem{gegenwart02}
P. Gegenwart {\em et al.},
Phys. Rev. Lett. {\bf 89}, 056402 (2002).


\bibitem{paschen04}
S. Paschen {\em et al.},
Nature {\bf 432}, 881 (2004).

\bibitem{friede10}
S. Friedemann {\em et al.},
PNAS {\bf 107}, 14547 (2010).

\bibitem{gegenwart06}
P. Gegenwart {\em et al.},
J. Phys. Soc. Jpn. {\bf 75} Suppl., 155 (2006).

\bibitem{gegenwart07}
P. Gegenwart {\em et al.},
Science {\bf 315}, 969 (2007).

\bibitem{oeschler08}
N. Oeschler {\em et al.},
Physica B {\bf 403}, 1254 (2008).

\bibitem{friede09}
S. Friedemann {\em et al.},
Nature Phys. {\bf 5}, 465 (2009).

\bibitem{custers10}
J. Custers {\em et al.},
\prl {\bf 104}, 186402 (2010).

\bibitem{kb_foot}
We employ units where $k_B=1$.

\bibitem{kuechler03}
R. K\"uchler {\em et al.},
Phys. Rev. Lett. {\bf 91}, 066405 (2003).

\bibitem{tokiwa09}
Y. Tokiwa {\em et al.},
Phys. Rev. Lett. {\bf 102}, 066401 (2009).

\bibitem{hartmann10}
S. Hartmann {\em et al.},
\prl {\bf 104}, 096401 (2010).

\bibitem{kusm08}
A link between Zeeman physics and $T^\ast$ was proposed before in
S. V. Kusminskiy {\em et al.},
Phys. Rev. B {\bf 77}, 094419 (2008),
but the relevant crossovers were not studied.

\bibitem{pw_esr}
A QP description has been successfully employed to model the
ESR spectra observed in \YRS\ \cite{pw09}.

\bibitem{pw09}
P. W\"olfle and E. Abrahams,
Phys. Rev. B {\bf 80}, 235112 (2009).


\bibitem{zwick_note}
Present-day ab-initio techniques cannot give reliable information on the
band structure on scales below 1 meV.

\bibitem{ssbook}
S.~Sachdev, {\it Quantum Phase Transitions},
Cambridge University Press, Cambridge (1999).

\bibitem{max_foot}
For a Zeeman-split local level, the maximum position in $\chi(T)$ is found from solving
$x n_F''(x) + n_F'(x)=0$ with $x=h/T$ and $n_F$ the Fermi function, resulting in $x =
1.54$. Similarly, the maximum in $\gamma(T)$ follows from $x n_F''(x) + 3 n_F'(x)=0$ with
the solution $x=3.24$.

\bibitem{scatt_note}
While the scattering rate from impurities will be affected by a Lifshitz
transition, those changes will cancel in $R_H$.

\bibitem{thall_note}
Different crossover criteria \cite{paschen04,friede10} may lead to $T_{\rm
Hall}$ values which differ by a factor of two for fixed $B$, due to the asymmetric
$R_H$ crossover [S. Friedemann, priv. comm.].


\bibitem{gegenwart05}
P. Gegenwart {\em et al.},
\prl {\bf 94}, 076402 (2005).


\end{thebibliography}
\end{document}